\documentclass{eas}
\usepackage[totalwidth=12.5cm, totalheight=19.0cm]{geometry}
\usepackage{graphicx}

\newcommand{\kms}{\hbox{km s$^{-1}$}}
\newcommand{\sauron}{\texttt{SAURON}}
\newcommand{\reffig}[1]{Fig.~\ref{#1}}

\begin{document}

\TitreGlobal{Mass Profiles and Shapes of Cosmological Structures}

\title{Dark Matter in the Central Regions of Early~Type Galaxies}

\author{Cappellari,~M.}\address{Leiden Observatory, Postbus 9513, 2300 RA Leiden, The Netherlands}
\author{Bacon,~R.}\address{Centre de Recherche Astrophysique de Lyon, Saint Genis Laval, France}
\author{Bureau,~M.}\address{Denys Wilkinson Building, University of Oxford, Keble Road, Oxford, United Kingdom}
\author{Damen,~M.~C.$^1$}
\author{Davies,~R.~L.$^3$}
\author{de Zeeuw,~P.~T.$^1$}
\author{Emsellem,~E.$^2$}
\author{Falc\'on-Barroso,~J.$^1$}
\author{Krajnovi\'c,~D.$^3$}
\author{Kuntschner,~H.}\address{Space Telescope European Coordinating Facility, European Southern Observatory, Germany}
\author{McDermid,~R.~M.$^1$}
\author{Peletier,~R.~F.}\address{Kapteyn Astronomical Institute, Postbus 800, 9700 AV Groningen, The Netherlands}
\author{Sarzi,~M.$^3$}
\author{van~den~Bosch,~R.~C.~E.$^1$}
\author{van~de~Ven,~G.$^1$}

\runningtitle{Dark Matter in the Central Regions of Early Type Galaxies}

\begin{abstract}
We investigate the well-known correlations between the dynamical mass-to-light ratio $M/L$ and other global observables of elliptical (E) and lenticular (S0) galaxies. We construct two-integral Jeans and three-integral Schwarzschild dynamical models for a sample of 25 E/S0 galaxies with \sauron\ integral-field stellar kinematics to about one effective (half-light) radius $R_{\rm e}$.  The comparison of the dynamical $M/L$ with the $(M/L)_{\rm pop}$ inferred from the analysis of the stellar population, indicates that dark matter in early-type galaxies contributes $\sim30\%$ of the total mass inside one $R_{\rm e}$, in agreement with previous studies, with significant variations from galaxy to galaxy. Our results suggest a variation in $M/L$ at constant $(M/L)_{\rm pop}$, which seems to be linked to the galaxy dynamics. We speculate that fast rotating galaxies have lower dark matter fractions than the slow rotating and generally more massive ones.
\end{abstract}
\maketitle

\section{Introduction}

\begin{figure}
\centering\includegraphics[height=6cm]{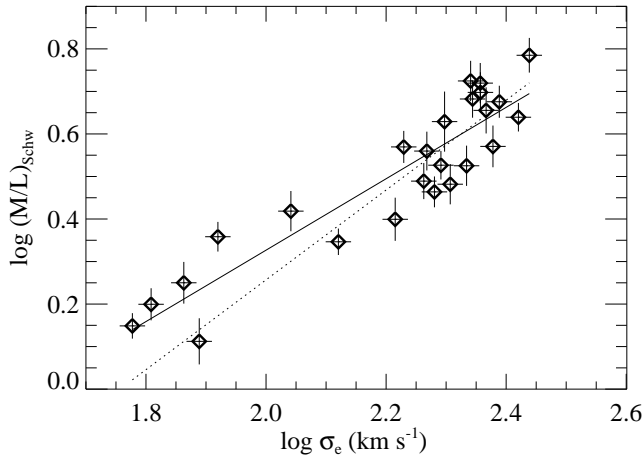}
    \caption{$(M/L)$--$\sigma_{\rm e}$ correlation. The solid line is the correlation $(M/L_I)=(3.80\pm0.14) \times (\sigma_{\rm e}/200\;\kms)^{0.84\pm0.07}$ obtained by fitting all galaxies, while the dotted line is the fit obtained by excluding galaxies with $\log\sigma_{\rm e}<2$. The galaxy with the smallest $\sigma_{\rm e}$ is M32.}
    \label{fig:ml-sigma}
\end{figure}

We measured the $M/L$ in the central regions, for a sample of 25 early-type galaxies with \sauron\ (Bacon et al.\ 2001) integral-field kinematics (Emsellem et al.\ 2004), using both two-integral Jeans models and three-integral Schwarzschild models (Cappellari et al.\ 2005). The galaxy sample was extracted from the \sauron\ survey (de Zeeuw et al.\ 2002) and spans over a factor of a hundred in mass. To this sample we added the galaxy M32 to explore the low-mass range. The models are constrained by \sauron\ kinematics, to about one effective (half-light) radius $R_{\rm e}$ and HST/WFPC2 $+$ MDM ground-based photometry in the $I$-band.  We studied the correlations of the $M/L$ with global observables of the galaxies and we found a sequence of increasing scatter when the $M/L$ is correlated to: (i) the luminosity-weighted second moment of the velocity inside the half-light radius $\sigma_{\rm e}$ (\reffig{fig:ml-sigma}), (ii) the galaxy mass, (iii) the $K$-band luminosity and (iv) the $I$-band luminosity. These correlations suggest that the $M/L$ depends primarily on galaxy mass, but at a given mass it is a function of $\sigma_{\rm e}$, which in turn is related to the galaxy concentration.  For our galaxy sample the $(M/L)$--$\sigma_{\rm e}$ relation has an observed scatter of 18\% and an inferred intrinsic scatter of $\sim13\%$.

\begin{figure}
  \centering\includegraphics[height=6cm]{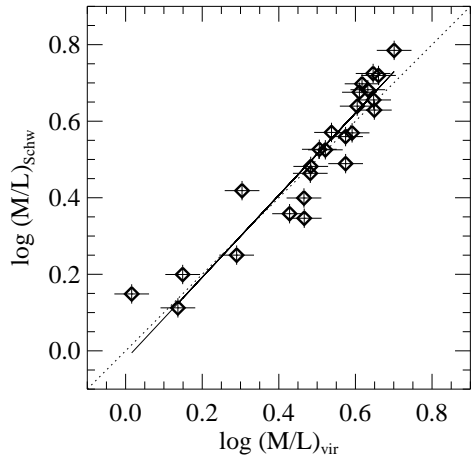}
    \caption{Comparison between the $(M/L)_{\rm vir}=\beta R_{\rm e}\sigma_{\rm e}^2/(L\, G)$ derived from the virial assumption and the $M/L$ obtained from the Schwarzschild models. The values of $(M/L)_{\rm vir}$ were scaled to match the dynamical $M/L$, and the best-fitting factor is $\beta=5.0\pm0.1$, in reasonable agreement with simple theoretical predictions. The solid line is a fit between the two quantities, while the dotted line represents a one-to-one correlation.}
    \label{fig:ml-schwarzschild-virial}
\end{figure}

We compared the $M/L$ derived from the dynamical models, which does not depend on any assumption of spatial or dynamical homology, with the classic predictions for the $(M/L)_{\rm FP}$ derived from the Fundamental Plane, which fully depends on the assumptions of homology and virial equilibrium (e.~g.\ Ciotti, Lanzoni \& Renzini 1996).  The slope of our observed correlations of $M/L$ with luminosity, mass and $\sigma_{\rm e}$, can account for 80--90\% of the tilt of the FP.  We also compared directly the $M/L$ from the models with the virial predictions, derived from integral-field kinematics, for our own sample (\reffig{fig:ml-schwarzschild-virial}), and found a relation $(M/L)\propto(M/L)_{\rm vir}^{1.08\pm0.07}$, which confirms, with smaller uncertainties, the results from the FP comparisons. This also shows that, when the effective radii are determined in the classic way using growth curves and $\sigma_{\rm e}$ is measured in an extended aperture as we do, the virial mass is a good estimator of galaxy mass. This has implications for high redshift studies, where the construction of full dynamical models is unfeasible, but the virial mass can still be mesured.

\section{Dark Matter Fraction}

\begin{figure}
	\centering\includegraphics[height=6cm]{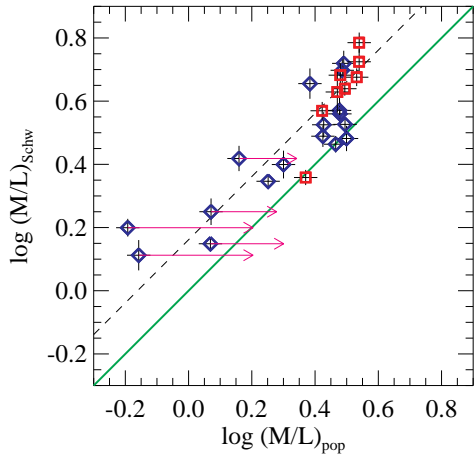}
	\caption{Dynamical (i.~e.\ total) $M/L$ from the Schwarzschild modeling as a function of $(M/L)_{\rm pop}$ using the Single Stellar Population models of Vazdekis et al.\ (1999), with a Kroupa (2001) IMF. The red squares and the blue diamonds indicate the slow and fast rotating galaxies respectively. The thick green line indicates the one-to-one relation. All galaxies have $(M/L)_{\rm pop} < (M/L)$ within the errors, but the total and stellar $M/L$ clearly do not follow a one-to-one relation. Dark matter is needed to explain the differences in $M/L$ (if the IMF is not varying).  The magenta arrows show the variation in the estimated $(M/L)_{\rm pop}$ for the youngest galaxies (luminosity-weighted age $<7$ Gyr), if a two-population model is assumed. The $(M/L)_{\rm pop}$ of the young galaxies would move closer to the one-to-one relation. Adopting the Salpeter (1955) IMF all the values of $(M/L)_{\rm pop}$ would increase by $\Delta\log(M/L)\sim 0.16$. This can be visualized by shifting the one-to-one relation to the position of the dashed line. In this case a number of galaxies would have $(M/L)_{\rm pop}>(M/L)$ and this implies that the Salpeter IMF is unphysical.}
	\label{fig:ml-mlpop}
\end{figure}

These correlations raise the question whether the observed $M/L$ variations are primarily due to differences in the $(M/L)_{\rm pop}$ of the stellar population or differences in the dark-matter fraction. To test this we compared our dynamical $M/L$ with simple estimates of the $(M/L)_{\rm pop}$, derived from the observed line-strengths indices of our galaxies (\reffig{fig:ml-mlpop}). We find that the $M/L$ broadly correlates with the $(M/L)_{\rm pop}$, indicating that the variation in the stellar population, mainly due to a variation in the luminosity-weighted age, is an important factor in driving the observed $M/L$ variation. The relatively small scatter in the correlation indicates that the IMF of the stellar population varies little among different galaxies, consistent with results obtained for spiral galaxies by Bell \& de Jong (2001).

The accuracy of our $M/L$ determinations allows us to detect significant deviations from a one-to-one correlation, which must be due to either variations of the IMF among galaxies, or to variations in the dark matter fraction within $R_{\rm e}$. In the latter case the dark matter fraction is found to be less than $\sim30\%$, consistent with earlier findings by e.~g.\ Gerhard et al.\ (2001). We find some evidence for the variation in $M/L$ to be related to the dynamics of the galaxies. In fact the slow rotating galaxies in our sample, which are more common among the most luminous objects, tend to have a higher $M/L$ at given $(M/L)_{\rm pop}$, than the fast rotating and generally fainter galaxies. Assuming a constant IMF among galaxies, these results would suggest that the slow rotating massive galaxies have a higher ($\sim30\%$) dark matter fraction than the fast rotating galaxies. We speculate that this difference in $M/L$ indicates a connection between the galaxy assembly history and the dark halo structure.


\vspace{-0.1cm}
\begin{thebibliography}{}

\bibitem{bac01}
Bacon R., et al.\ 2001, MNRAS, 326, 23

\bibitem{bel01}
Bell, E.~F., de Jong, R.~S.\ 2001, ApJ, 550, 212

\bibitem{cap05} Cappellari, M., et al.\ 2005, MNRAS, in press (astro-ph/0505042)

\bibitem{1996MNRAS.282....1C} Ciotti, L., Lanzoni, B., Renzini, A.\ 1996, MNRAS, 282, 1

\bibitem{dez02} de Zeeuw, P.~T., et al.\ 2002,
MNRAS, 329, 513

\bibitem{ems04}
Emsellem, E., et al.\ 2004, MNRAS, 352, 721

\bibitem{2001AJ....121.1936G} Gerhard, O., Kronawitter, A., Saglia, R.~P.,
Bender, R.\ 2001, AJ, 121, 1936

\bibitem{kro01} Kroupa, P.\ 2001, MNRAS, 322, 231

\bibitem{1955ApJ...121..161S}
Salpeter, E.~E.\ 1955, ApJ, 121, 161

\bibitem{1999ApJ...513..224V}
Vazdekis, A.\ 1999, ApJ, 513, 224

\end{thebibliography}
\end{document}